\documentclass[aps,prd,nofootinbib,twocolumn,showpacs]{revtex4-1}
\usepackage{dcolumn}
\usepackage{lipsum}
\usepackage{graphicx,epsfig}
\usepackage{psfrag}
\usepackage{bm}
\usepackage{epstopdf}
\usepackage{latexsym}
\usepackage{multirow}
\usepackage{amsmath,amssymb}
\usepackage{enumerate}
\usepackage{hyperref}
\usepackage[FIGTOPCAP]{subfigure}

\def\be {\begin{equation}}
\def\ee {\end{equation}}
\def\ba {\begin{eqnarray}}
\def\ea {\end{eqnarray}}
\def\nn {\nonumber}
\def\bc {\begin{center}}
\def\ec {\end{center}}
\newcommand{\bdm}{\begin{displaymath}}
\newcommand{\edm}{\end{displaymath}}

\def\a  {\alpha}
\def\b  {\beta}
\def\g  {\gamma}
\def\G  {\Gamma}
\def\d  {\delta}

\def\l  {\lambda}
\def\m  {\mu}
\def\n  {\nu}

\def\O  {\Omega}

\def\r  {\rho}

\def\s {\sigma}

\def\t  {\tau}

\def\nn {\nonumber}
\def\ra {\rightarrow}
\def\na {\nabla}
\def\da {\dagger}
\def\la {\label}
\def\le {\left}
\def\ri {\right}
\def\pa {\partial}
\def\f {\frac}
\def\sq {\sqrt}

\def\bi {\begin{itemize}}
\def\ei {\end{itemize}}

\def\> {\rangle}
\def\< {\langle}

\def\bc {\begin{center}}
\def\ec {\end{center}}

\begin{document}


\title{Spin $1/2$ field and regularization in de Sitter and radiation dominated universe}

\author{Suman Ghosh} \email[email: ]{smnphy@gmail.com}
\affiliation{Tata Institute of Fundamental Research, Homi Bhabha Road, Mumbai - 400005, India}

%

\begin{abstract}

We construct a simple algorithm to derive number density of spin $1/2$ particles created in spatially flat FLRW spacetimes and resulting renormalized energy-momentum tensor within the framework of adiabatic regularization. 
Physical quantities thus found are in agreement with the known results. This formalism can be considered as an appropriate extension of the techniques originally introduced for scalar fields, applicable to fermions in curved space. We apply this formalism to compute the particle number density and the renormalized energy density and pressure analytically (wherever possible) and numerically, in two interesting cosmological scenarios: a de Sitter spacetime and a radiation dominated universe. Results prove the efficiency of the methodology presented here.
\end{abstract}

\pacs{04.62.+v, 11.10.Gh 	}
\maketitle

\section{Introduction}

One of the inevitable quantum consequences of the dynamical evolution of the universe is particle creation by quantum fields in a classical gravitational background \cite{PT,MW,Wald-QFTCS,Buchbinder,Fulling,BD}. Such phenomenon was first discovered by Parker \cite{Parker-thesis, Parker:1968mv, Parker:1971pt}. This is conceptually similar to the black hole radiation \cite{Hawking:1974sw}. Particle creation during very early moments of the universe is useful in explaining the inhomogeneities in the cosmic microwave background (CMB) radiation and the large-scale structure of the Universe \cite{Liddle:2000cg}. Recently attempts have been made to quantify effects of renormalization on the correlation functions visible in CMB power spectrum \cite{Agullo:2009vq, Agullo:2008ka}. Concept of adiabatic vacuum is crucial to have a notion of particles in curved space that comes closest to that in flat space. 
Studies on quantum scalar fields in curved space-time in spatially flat Friedmann-Lemaitre-Robertson-Walker (FLRW) spacetimes revealed the following generic features -- (i) in a comoving volume, the particle number density ($|\b_k|^2$) in mode `$k$' is an adiabatic invariant (which corresponds to conservation of number of quanta produced by a quantum oscillator whose length is decreasing infinitely slowly), (ii) particles of conformally invariant field with zero mass will not be created, (iii) the total number density of particles of specific mass, summed over modes is ultraviolet (UV) divergent and (iv) the resulting stress tensor ($T_{\m\n}$) has quadratic and logarithmic UV  divergences in addition to the usual quartic divergence or the vacuum energy in flat space.

{\em Adiabatic regularization} was introduced by Parker \cite{parker:2012} to make the total number density for scalar particles finite and later extended to tame the UV divergences in $T_{\m\n}$ by Parker and Fulling \cite{Parker:1974qw}.
Adiabatic regularization of formal quantities having divergences is achieved  by subtracting mode by mode (under the integral sign) each term in the adiabatic (large mode) expansion of the integrand that contains at least one UV divergent part for arbitrary values of the parameters of the theory. The number of time derivatives of the scale factor in a term of the expansion is called the {\em adiabatic order} of that term. This scheme is useful for numerical calculations.
In \cite{Parker:1974qw}, the authors have shown that the adiabatic regularization is equivalent to the $n$-wave regularization (which is  a variant of Pauli-Villars regularization \cite{BD}) used by Zeldovich and Starobinsky \cite{Zeldovich:1971mw}. Later method is also useful for numerical computations and used successfully \cite{Garriga:1989jx, Ghosh:2008zs} in cases where it is difficult to find exact solution to the field equation due to complexity of the metric functions.

There are other powerful regularization methods, that were developed to apply in curved space, such as proper-time regularization, point-splitting regularization (particularly by Hadamard method), zeta-function regularization, dimensional regularization etc \cite{PT,MW,Wald-QFTCS,Fulling,BD}. 
Recently, the DeWitt-Schwinger point-splitting regularization \cite{DeWitt:1975ys, Christensen:1976vb, Birrell-1978, Anderson:1987yt, Christensen:1978yd}, is used in \cite{Matyjasek:2013vwa, Matyjasek:2014lja} in the context of massive scalar, spinor and vector fields in spatially flat FLRW universe using asymptotic expansion of the Green function \cite{Zeldovich:1971mw, Beilin:1980es} to reproduce the leading order contribution to the stress tensor found in \cite{Kaya:2011yu}. All these methods are equivalent \cite{BD}. 
The renormalised quantities in this methods are mostly computed directly from the action and not via solving for $|\b_k|^2$. 
Production of spin $1/2$ particles in various cosmological scenarios have been studied by many \cite{Unruh:1974bw, Campos:1991ff, Gibbons:1993hg, Villalba:1995za, Brustein:2000hi, Anischenko:2009va} over the years.
Recently, fundamental issues like definition of a preferred vacuum state at a given time \cite{Agullo:2014ica, Anderson:2005hi} and
approximate definition of the particle number via adiabatic WKB ansatz \cite{Winitzki:2005rw}  is addressed.
A systematic adiabatic expansion in FLRW universe for spin $1/2$ field modes has been  constructed in \cite{Landete:2013axa, Landete:2013lpa, delRio:2014cha} and corresponding renormalization is achieved. This lead to a proof of the equivalence between the adiabatic regularization and the DeWitt-Schwinger method \cite{delRio:2014bpa}. It was argued in \cite{Landete:2013axa} that the WKB ansatz is specifically designed to preserve the Klein-Gordon product and the associated Wronskian condition, but is not useful to preserve the Dirac product and the associated Wronskian. In other words, it is not possible to extend the WKB ansatz for the fermionic modes beyond the zeroth adiabatic order.

In \cite{Ghosh:2015mva}, we introduced a formalism, alternative to \cite{Landete:2013axa}, to determine $|\b_k|^2$ and also to regularize the resulting $T_{\m\n}$ for spin $1/2$ particles in spatially flat FLRW universe  within the framework of adiabatic regularization. This methodology can also be regarded as an extension of the formalism presented in \cite{Zeldovich:1971mw} (in the context of scalar particle creation during anisotropic collapse) applicable to spin $1/2$ fields.
In the following, we review and add to the formalism presented in \cite{Ghosh:2015mva} and present illustrative examples of application to cosmological scenarios.
We define our `in' state as the adiabatic vacuum (state of adiabatic order zero) given by the WKB solution to the field equations and the adiabatic matching with exact state is done using via {\em time-dependent} Bogolubov coefficients. Next we use the field equations to derive the governing equations for $|\b_k|^2$, in terms of a set of three {\em real} and independent variables $s_k, u_k$ and $\t_k$ (to be defined later), that were introduced in \cite{Zeldovich:1971mw}. 
It is then easy to regularize $T_{\m\n}$ by subtracting leading order terms from the adiabatic expansions of these variables. The renormalized quantities and conformal anomaly thus derived match exactly with the known results \cite{BD, PT, Landete:2013axa}. We also show that this method is manifestly covariant i.e. the resulting renormalized stress tensor is conserved.
Then we apply this formalism to two cosmological scenarios of interest-- de Sitter and radiation dominated universes. We derived analytic (wherever possible) and numerical results (presented via plots) that matches the already known results to prove the validity and effectiveness of the formalism. 

\section{Dirac field in FLRW spacetime}  

The homogeneous and isotropic FLRW spacetime geometry is given by
\be
ds^2 = - dt^2 + a^2(t)\,d\vec{x}^2, \la{eq:metric1}
\ee
where $t$ is the cosmological time and $a(t)$ is the cosmological scale factor. Now to appreciate the conformal symmetry of such background we move to conformal time ($\eta = \int \f{dt}{a(t)}$) description of the metric,
\be
ds^2 =  a^2(\eta)(- d\eta^2 + d\vec{x}^2). \la{eq:metric2}
\ee
Dirac equation in generic curved spacetime for field $\psi(\vec{x},\eta)$ with mass $m$ is given by \cite{PT,MW,Wald-QFTCS,Fulling,BD},
\be
(e^\m_a\g^a \nabla_\m - m)\psi = 0. \la{eq:Dirac1}
\ee
where $e^a_\m$ are the vierbeins. Greek alphabets denote coordinate indices and Latin alphabets denote frame indices. $\g^a$'s are Dirac matrices (defined in terms of usual Pauli matrices $\s^i$) in Minkowski space, satisfying $\{\g^a,\g^b\} = \eta^{ab}$ where $\eta^{ab}$ is the Minkowski metric $(-1,1,1,1)$ and $\nabla_\m = \pa_\m - \G_\m$ is the covariant derivative where $\G_\m$ is the spin connection \cite{PT,BD}. 
The Dirac matrices in the Dirac-Pauli representation is given by
\ba
\g^0 = i \le( \begin{array}{cc} 
I & 0 \\
0 & -I \\ 
\end{array} \ri), ~~~~~ \g^i = i\le( \begin{array}{cc} 
0 & \s^i   \\
-\s^i & 0 \\ 
\end{array} \ri)
\ea
where $\s^i$ are the Pauli matrices,
\ba
\s^1 = \le( \begin{array}{cc} 
0 & 1   \\
1 & 0 \\ 
\end{array} \ri),~~\s^2 = \le( \begin{array}{cc} 
0 & -i   \\
i & 0 \\ 
\end{array} \ri),~~\s^3 = \le( \begin{array}{cc} 
1 & 0   \\
0 & -1 \\ 
\end{array} \ri).
\ea
For metric (\ref{eq:metric2}), Eq. (\ref{eq:Dirac1}) leads to
\be
\le[\g^0 \le(\pa_0 + \f{3\dot{a}}{2a}\ri) + \g^i \pa_i + ma\ri] \psi = 0 \la{eq:Dirac2}
\ee	
where a over-dot represent derivative with respect to $\eta$. Dirac field $\psi$ can be written in terms of {\em time-dependent} annihilation operator for particles ($B_{\vec{k}\l}(\eta)$) and creation operator for antiparticles ($D^{\da}_{\vec{k}\l}(\eta)$) as
\be
\psi = \sum_{\l} \int d^3k \big(B_{\vec{k}\l}u_{\vec{k}\l} + D^{\da}_{\vec{k}\l}v_{\vec{k}\l}\big) \la{eq:psi-gen}
\ee
where mode expansion of the eigenfunctions $u_{\vec{k}\l}(\vec{x},\eta)$ and $v_{\vec{k}\l}(\vec{x},\eta)$, which is obtained by charge conjugation ($v = \g^2u^*$) operation on $u_{\vec{k}\l}(\vec{x},\eta)$, are given, in terms of two component spinors, by \cite{Landete:2013axa}
\ba
u_{\vec{k}\l}(\vec{x},\eta) &=& \f{e^{{i \vec{k}.\vec{x}}}}{(2\pi a)^{3/2}} \le( \begin{array}{l} 
h^{I}_{k}(\eta) \xi_\l(k)  \\
h^{II}_{k}(\eta) \f{\vec{\s}.\vec{k}}{k} \xi_\l(k) \\ 
\end{array} \ri)  \la{eq:+psi_ansatz} \\
v_{\vec{k}\l}(\vec{x},\eta) &=& \f{e^{{-i \vec{k}.\vec{x}}}}{(2\pi a)^{3/2}} \le( \begin{array}{l} 
-h^{II*}_{k}(\eta) \f{\vec{\s}.\vec{k}}{k} \xi_{-\l}(k)  \\
-h^{I*}_{k}(\eta)  \xi_{-\l}(k) \\ 
\end{array} \ri)  \la{eq:-psi_ansatz}
\ea 
where $\xi_\l(k)$ is the normalised two component spinor satisfying $\xi_\l^\da \xi_\l = 1$ and $\f{\vec{\s}.\vec{k}}{2k}\xi_\l = \l\xi_\l$ where $\l = \pm 1/2$ represents the helicity\footnote{Using either value of helicity or either of $u_{\vec{k}\l}$ and $v_{\vec{k}\l}$, in the following calculations, leads to exactly same end results.}. Normalisation condition in terms of Dirac product for $\psi$, $(u_{\vec{k}\l}, u_{\vec{k'}\l'}) = (v_{\vec{k}\l}, v_{\vec{k'}\l'}) = \d_{\l\l'}\d(\vec{k}-\vec{k'})$, implies
\be
|h^{I}_{k}(\eta)|^2 + |h^{II}_{k}(\eta)|^2 = 1. \la{eq:norm_h}
\ee
This condition ensures that the standard anti-commutation relations for creation and annihilation operators are satisfied.
Putting Eq. (\ref{eq:+psi_ansatz}) in Eq. (\ref{eq:Dirac2}), we get the following first order coupled differential equations
\ba 
{\dot h}^{I}_k + i m a\, h^{I}_{k} + i k\, h^{II}_{k} &=& 0,  \la{eq:h1h2}\\
{\dot h}^{II}_k - i m a\, h^{II}_{k} + i k\, h^{I}_{k} &=& 0. \la{eq:h2h1}
\ea
with the Wronskian 
\be
{\dot h}^{I}_k\, h^{II*}_k - h^{I}_k\,{\dot h}^{II*}_k = -i k. \la{eq:wronskian1} 
\ee
Eq. (\ref{eq:h1h2}) and (\ref{eq:h2h1}) can be decoupled to give the following decoupled second order equations
\ba
{\ddot h}^I_k + \big[\O_k^2(\eta) + i Q(\eta)\big] h^{I}_{k} &=& 0, \la{eq:h1} \\
{\ddot h}^{II}_k + \big[\O_k^2(\eta) - i Q(\eta)\big] h^{II}_{k} &=& 0, \la{eq:h2}
\ea
where $\O_k(\eta) = \sq{m^2a^2 + k^2}$ and $Q(\eta) = m \dot{a}$. 
The adiabatic vacuum (i.e. state of adiabatic order zero) is given by the WKB solution of the field equations (see Appendix \ref{app-1}) that naturally generalise the standard Minkowski space solution,
\ba
h^{I(0)}_k(\eta)  = f_1\, e_-,~~~~ h^{II(0)}_k(\eta)  = f_2\, e_- \la{eq:g1g2}
\ea
with 
\be
f_1 = \sq{\f{\O_k + ma}{2\O_k}}, f_2 = \sq{\f{\O_k - ma}{2\O_k}}, e_{\pm} = e^{\pm i\int \O_k d\eta}. \la{eq:f1f2e}
\ee
This implies the exact solution can be written as 
\ba
h^{I}_k(\eta) &=& \a_k(\eta) h^{I(0)} - \b_k(\eta) h^{II(0)*}, \la{eq:h1ansatz}\\
h^{II}_k(\eta) &=& \a_k(\eta) h^{II(0)} + \b_k(\eta) h^{I(0)*} \la{eq:h2ansatz}
\ea
where $\a_k(\eta)$ and $\b_k(\eta)$ are the Bogolubov coefficients. The normalisation condition (\ref{eq:norm_h}) leads to
\be
|\a_k(\eta)|^2 + |\b_k(\eta)|^2 = 1. \la{eq:norm_ab}
\ee
Here the adiabatic matching \cite{BD} is used to obtain a zeroth order adiabatic vacuum state at a particular initial time, say $\eta = \eta_0$. This matching fixes the initial values of the mode functions and their first derivatives at $\eta_0$. Note that, adiabatic matching at different times results in different states for the field. In this case, Eqs. (\ref{eq:h1ansatz}) and (\ref{eq:h2ansatz}) implies, the initial conditions are $\a_k(\eta_0)=1$ and $\b_k(\eta_0)=0$. Here, $\a_k(\eta)$ and $\b_k(\eta)$ carries all the {\em non-adiabaticity} (see the adiabatic expansion of $|\b_k(\eta)|^2$ shown later). If one chooses a state of different adiabatic order, for matching with the exact state, the associated Bogolubov coefficients (and their initial values at $\eta_0$) will be different. 
In general, one can write \cite{Landete:2013axa} 
\ba
h^{I}_k(t) &=& \a_k^{(n)}(\eta) h^{I(n)} - \b_k^{(n)}(\eta) h^{II(n)*}, \la{eq:h1nansatz}\\
h^{II}_k(\eta) &=& \a_k^{(n)}(\eta) h^{II(n)} + \b_k^{(n)}(\eta) h^{I(n)*} \la{eq:h2nansatz}
\ea
where $h^{I(n)}$ and $h^{II(n)}$ are solutions of $n^{th}$ adiabatic order and $\a_k^{(n)}$ and $\b_k^{(n)}$ carries terms of adiabatic order higher than `$n$' i.e. they are constants upto adiabatic order `$n$'. However, as it is impossible to find consistent WKB solution of higher order for spin $1/2$ fields, we consider the matching with zeroth adiabatic order only. 
Our ansatz, Eqs. (\ref{eq:h1ansatz}) and (\ref{eq:h2ansatz}) further implies 
\ba
\a_k(\eta) &=& \Big(f_1\, h^{I}_k + f_2\, h^{II}_k\Big)e_+,  \la{eq:alpha1} \\
\b_k(\eta) &=& \Big(f_1\, h^{II}_k - f_2\, h^{I}_k\Big)e_-. \la{eq:beta1}
\ea 
The average number density of spin $1/2$ particles of specific helicity and charge with momentum $\vec{k}$ is then \cite{PT},
\be
\langle N_{\vec{k}}\rangle = \langle B^{\da}_{\vec{k}\l} B_{\vec{k}\l}\rangle = \langle D^{\da}_{\vec{k}\l} D_{\vec{k}\l}\rangle = |\b_k(\eta)|^2. \la{eq:no.den.}
\ee
Note that, the WKB solutions (\ref{eq:g1g2}) obey the following Wronskian condition
\be
{\dot h}^{I(0)}_k\, h^{II(0)*}_k - h^{I(0)}_k\,{\dot h}^{II(0)*}_k = F - i k, ~~~~F = \f{k\,Q}{2\O_k^2}. \la{eq:wronskian2} 
\ee
The function $F(\eta)$ is of adiabatic order one. 
In the adiabatic limit ($F\ra 0$), Eq. (\ref{eq:g1g2}) do satisfy the exact Wronskian 
(\ref{eq:wronskian1}) and thus represents the adiabatic vacuum.
Eq. (\ref{eq:wronskian2}) also implies that the WKB ansatz is not an exact solution (due to presence of $F(\eta)$) of the field equation during the (non-adiabatic) expansion which is the desired condition for particle creation \cite{parker:2012}. Thus $F(\eta)$ is a measure of non-adiabaticity of the cosmological evolution.
In Eq. (\ref{eq:psi-gen}), the creation and annihilation operators carry this non-adiabaticity and so as the Bogolubov coefficients in Eqs (\ref{eq:h1ansatz} - \ref{eq:h2ansatz}). Thus $|\b_k|^2$ is expected to depend on $F(\eta)$ as particle creation can be considered as a quantum consequence of this non-adiabaticity.
Putting Eqs (\ref{eq:h1ansatz}) and (\ref{eq:h2ansatz}) in Eqs (\ref{eq:h1h2}) and (\ref{eq:h2h1}), a system of coupled linear first order differential equations is obtained for $\a_k(\eta)$ and $\b_k(\eta)$:
\be
\dot{\a}_k = -F\, \b_k\, e_+^2, ~~~~~~\dot{\b}_k = F\, \a_k\, e_-^2. \la{eq:alphabeta}
\ee
Eq. (\ref{eq:alphabeta}) implies the usual absence of massless particles in conformally flat spacetimes.
Further, fermions at rest will also not be produced. Similar results were found in \cite{Khanal:2013cjp} using Newman-Penrose formalism. 
To determine $|\b_k|^2$ and the resulting $T_{\m\n}$, we define  three real and independent variables $s_k$,  $u_k$ and $\t_k$ (following \cite{Zeldovich:1971mw}), in terms of the two complex variables $\a_k$ and $\b_k$ as,
\ba
s_k = |\b_k |^2,~~ u_k = \a_k\,\b_k^*\,e_-^2 + \a_k^*\,\b_k\,e_+^2, \nn \\
\t_k = i(\a_k\,\b_k^*\,e_-^2 - \a_k^*\,\b_k\,e_+^2). \la{eq:def_sut}  
\ea
For these variables one gets a system of three linear first order differential equations:
\ba
\dot{s}_k &=& F\ u_k, \la{eq:s}\\
\dot{u}_k &=& 2F(1-2s_k) - 2 \O_k \t_k, \la{eq:u}\\
\dot{\t}_k &=& 2\O_k u_k. \la{eq:t}
\ea
with initial conditions $s_k = u_k = \t_k = 0$ at suitably chosen time $\eta= \eta_0$ as discussed earlier. 
Eqs (\ref{eq:s}-\ref{eq:t}) are the key equations of this formalism.
Note that, if one can derive $h^{I}_k$ and $h^{II}_k$ analytically from the field equations, then $|\b_k|^2$ directly follows from Eq. (\ref{eq:beta1}). Alternatively, one can solve the set of equations (\ref{eq:s}-\ref{eq:t}) numerically (or analytically whenever possible). 

\subsection{Energy-momentum tensor}

The energy-momentum tensor for Dirac field in curved spacetime is given by
\be
T_{\m\n} = \f{i}{2} \Big[\bar{\psi}\g_{(\m}\na_{\n)}\psi - (\na_{(\m}\bar{\psi})\g_{\n)}\psi\Big] \la{eq:stress}
\ee
The independent components of $T_{\m\n}$, the energy density $T^0_0$ and pressure density $T^i_i$ are
\ba
T^0_0 &=& -\f{i}{2a} \Big(\bar{\psi}\g^0\dot{\psi} - \dot{\bar{\psi}}\g^0\psi\Big), \la{eq:T00}\\
T^i_i &=& \f{i}{2a} \Big(\bar{\psi}\g^i\psi' - \bar{\psi}'\g^i\psi\Big), \la{eq:Tii}
\ea
where ($'$) denotes derivative with respect to $x^i$. Using eqs (\ref{eq:psi-gen}-\ref{eq:-psi_ansatz}), vacuum expectation value of the above quantities can be written as
\ba
\< T^0_0 \> \equiv \r &=& \f{1}{(2\pi a)^3} \int d^3k \, \r_k, \la{eq:<T00>}\\
\< T^i_i \> \equiv p &=& \f{1}{(2\pi a)^3} \int d^3k \, p_k, \la{eq:<Tii>}
\ea
where the energy density $\rho_k$ and the pressure density $p_k$, for each mode, are given respectively as,
\ba
\r_k &=& \f{i}{a} \Big(h^{I}_k \dot{h}^{I*}_k + h^{II}_k \dot{h}^{II*}_k - h^{I*}_k \dot{h}^{I}_k - h^{II*}_k \dot{h}^{II}_k\Big), \la{eq:r_k} \\
p_k &=&  \f{2k}{3a} \Big(h^{I}_k h^{II*}_k + h^{I*}_k h^{II}_k\Big). \la{eq:p_k}
\ea
Using Eqs (\ref{eq:h1ansatz}), (\ref{eq:h2ansatz}), (\ref{eq:norm_ab}) and (\ref{eq:def_sut}), we get from Eqs (\ref{eq:r_k}) and (\ref{eq:p_k})
\ba
\r_k &=& -\f{2\O_k}{a}\big(1-2s_k\big), \la{eq:rho_k_sut} \\
p_k &=& \f{2k}{3a} \le[\f{k}{\O_k}(1- 2s_k) + \f{ma}{\O_k} u_k\ri]. \la{eq:p_k_sut}
\ea
The vacuum energy (i.e. in absence of any gravitationally created particles when $s_k = u_k = \t_k = 0$) matches with the standard result. 
We discuss below how to remove the divergences in $T_{\m\n}$ by subtracting the leading order terms from the adiabatic expansion of $s_k, u_k, \t_k$.


\subsection{Renormalization}

One can expand the solutions of the system of Eqs (\ref{eq:s}-\ref{eq:t}) in an asymptotic series in powers of $\O_k^{-1}$ in the large momenta limit ($\O_k \ra \infty$). This is essentially same as the adiabatic expansion that is valid in the quasi-classical region where $|\dot{\O}_k|<< \O_k^2$. The adiabatic expansions of $s_k, u_k, \t_k$ gives (Appendix \ref{app-2}), $\t_k = \t_k^{(1)} + \t_k^{(3)} + ...$, $u_k = u_k^{(2)} + u_k^{(4)} + ...$ and $s_k = s_k^{(2)} + s_k^{(4)} + ...$, where the superscripts inside the brackets indicates the adiabatic order.
Eqs (\ref{eq:s}-\ref{eq:t}) leads to the following recursion relations among $s_k^{(r)}, u_k^{(r)}$ and $\t_k^{(r)}$, 
\ba
u_k^{(r)} &=& \f{\dot{\t}_k^{(r-1)}}{2\O_k}, \la{eq:u^r}\\
s_k^{(r)} &=& \int F u_k^{(r)} dt, \la{eq:s^r}\\
\t_k^{(r+1)} &=& -\f{4F s_k^{(r)} + \dot{u}_k^{(r)}}{2\O_k}, \la{eq:t^r}
\ea
with $r = 2,4,...$ and $\t_k^{(1)} = \f{F}{\O_k}$. It is straightforward to solve these equations analytically to arbitrary order. Further, as $k \ra \infty$ and if derivatives of $a(\eta)$ to any order is non zero, we have
\be
s_k^{(r)} \sim k^{-(r+2)},~~ u_k^{(r)} \sim k^{-(r+1)}, ~~ \t_k^{(r)} \sim k^{-(r+1)}. \la{eq:lim_sut}
\ee
The leading order terms imply the well-known logarithmic UV divergences of the total energy and pressure density. Note that there is no quadratic divergence that appears for scalar fields \cite{PT}. Further, as $k \ra \infty$, the leading term in particle number density $s_k$ decays as $k^{-4}$ irrespective of the functional form of the scale factor. To remove the infinities, we need to subtract leading terms upto necessary order from the expansion of $s_k$, $u_k$ and $\t_k$. 
Note that the total particle number density of specific mass with summed over momenta is given by
\be 
N_m = \f{1}{(2\pi a)^3} \int d^3k \,s_k. 
\la{eq:totalPND}
\ee
and is finite unlike the energy and pressure densities.
The renormalized total energy and momentum density (after subtracting the vacuum contribution i.e. the quartic divergence) are given by

\ba
\r_{{}_{Ren}} &=& \f{2}{\pi^2 a^4} \int dk\, k^2\, \O_k \Big(s_k - s_k^{(2)} - s_k^{(4)}\Big), \la{eq:r_k_ren_gen} \\
p_{{}_{Ren}} &=&  \f{1}{3\pi^2 a^4} \int dk \f{k^3}{\O_k} \Big[-2k \big(s_k  - s_k^{(2)} - s_k^{(4)}\big) \nn \\
&& \hspace{1.5cm} + ma \big(u_k  - u_k^{(2)} - u_k^{(4)}\big)\Big]. \la{eq:p_k_ren_gen}
\ea
Note that, in the specific case of FLRW background, it is enough to subtract only upto second adiabatic order. However, in more generic spacetimes the fourth-order adiabatic terms may give rise to proper UV divergences \cite{Christensen:1978yd} and therefore the standard approach of regularization considers the fourth order adiabatic terms as {\em potentially divergent} \cite{PT,BD}.

\subsection{Conformal anomaly}
The trace of the energy momentum tensor (\ref{eq:stress}), $T^\m_\m = m \bar{\psi} \psi$, 
vanishes for massless fields. However, renormalization procedure renders the quantum counterpart of $T^\m_\m$ finite. This phenomenon is known as the conformal anomaly. 
The vacuum expectation value of the trace of stress tensor is
\be
\< T^\m_\m \> = \f{1}{(2\pi a)^3} \int d^3k \, \< T^\m_\m \> _k \la{eq:<Tmm>}
\ee
with
\ba
\< T^\m_\m \> _k &=&  -2m\Big(|h^{I}_k|^2 - |h^{II}_k|^2\Big) \la{eq:<Tmm>k} \\
&=& - 2m \le[\f{ma}{\O_k}(1-2s_k) - \f{k}{\O_k}u_k\ri] \la{eq:<Tmm>k_sut}
\ea
using  Eqs (\ref{eq:h1ansatz}), (\ref{eq:h2ansatz}), (\ref{eq:norm_ab}) and (\ref{eq:def_sut}). Alternatively, Eq. (\ref{eq:<Tmm>k_sut}) follows from the identity $\< T^\m_\m \> _k = \r_k + 3p_k$ too.
In the limit $m\ra 0$, subtracting the vacuum contribution and terms upto fourth adiabatic order, the resulting renormalized trace anomaly is given by,
\ba
\< T^\m_\m \> _{Ren} &=& \lim_{m\ra 0} \f{2m}{(2\pi a)^3} \int d^3k \le[\f{2ma}{\O_k}\le(s_k - s_k^{(2)} - s_k^{(4)}\ri)\ri. \nn\\ 
&& ~~~~~~~~~~~~~~~\le. + \f{k}{\O_k} \le(u_k - u_k^{(2)} - u_k^{(4)}\ri)\ri]\\
&=&- \lim_{m\ra 0} \f{2m}{(2\pi a)^3} \int d^3k \le[\f{2ma}{\O_k} s_k^{(4)} + \f{k}{\O_k} u_k^{(4)}\ri]~~~~ \la{eq:<Tmm>ren1}
\ea
as only the fourth order term in the expansions of $s_k(\eta)$ and $u_k(\eta)$ survives in the $m\ra 0$ limit.	
Using explicit expressions of $s_k^{(4)}$ and $u_k^{(4)}$ (Appendix \ref{app-2}) in Eq. (\ref{eq:<Tmm>ren1}), we get\footnote{Here we correct an overall sign mistake made in \cite{Ghosh:2015mva}.}
\be
\< T^\m_\m \> _{Ren} = - \f{11\dot{a}^4 - 29 a \dot{a}^2\ddot{a} + 12  a^2\dot{a}\dddot{a} + 9 a^2 \ddot{a}^2 - 3 a^3\ddddot{a}}{240\pi^2 a^8}. \la{eq:<Tmm>ren2}
\ee
Note that the conformal anomaly can be expressed in terms of the curvature invariants as \cite{PT,BD},
\be
\< T^\m_\m \> _{Ren} = \f{1}{(4\pi)^2} \le(A\, C_{\a\b\g\d}C^{\a\b\g\d} + B\, G + C\, \square R \ri), \la{eq:Tmm_gen}
\ee
where $C_{\a\b\g\d}$ is the Weyl tensor, $R$ is the Ricci scalar and $G$ is the Gauss-Bonnet invariant, given by $G = -2(R_{\a\b}R^{\a\b} - R^2/3)$ with $R_{\a\b}$ being the Ricci tensor. For conformally flat spacetimes, Weyl tensor vanishes identically. Equating Eq. (\ref{eq:Tmm_gen}) with Eq. (\ref{eq:<Tmm>ren2}), we get $B = 11/360$ and $C=-1/30$, which agrees with the known results \cite{PT,BD}. 
Similarly one can check that the axial anomaly vanishes as expected \cite{Ghosh:2015mva}.

\subsection{Conservation of stress tensor}

The renormalized energy momentum tensor constructed here is manifestly a conserved quantity. This can be checked explicitly. It is easy to see that the conservation condition
\be
\< T^{i\n}_{;\n} \> = 0, ~~ i = 1,2,3
\ee
is satisfied identically and the remaining component of the stress tensor conservation equation
\ba
\< T^{0\n}_{;\n} \> = \< \dot{T}^{00} \> + 5\f{\dot{a}}{a}\< T^{00} \> + 3\f{\dot{a}}{a} \< T^{ii} \> .
\ea
vanishes identically by virtue of Eqs (\ref{eq:r_k_ren_gen}), (\ref{eq:p_k_ren_gen}) and (\ref{eq:s}). 

\section{Application to de Sitter and radiation dominated universe}

Below we apply the formalism described above to two specific examples: de Sitter spacetime and radiation dominated universe.

\subsection{de Sitter universe}

The de Sitter universe is characterized by the scale factor $a(t) = e^{Ht}$ where $H$ is the Hubble constant. In terms of conformal time we have $a(\eta) = -1/H\eta, -\infty < \eta < 0$.
Solutions to Eq. (\ref{eq:h1}) and (\ref{eq:h2}), in a de Sitter universe that has the correct asymptotic behaviour is given by \cite{delRio:2014cha}
\ba
h^I(z) &=& i \f{\sq{\pi z}}{2} e^{\pi \m/2} H^{(1)}_{\n_+}(z) ,  \la{eq:h1_dS}\\
h^{II}(z) &=& \f{\sq{\pi z}}{2} e^{\pi \m/2} H^{(1)}_{\n_-}(z)  \la{eq:h2_dS}
\ea
where $z=-k\eta$, $\m = m/H$, $\n_{\pm} = \pm \f{1}{2} - i \m$ and $H^{(1)}_{\n_\pm}(z)$  are the Hankel functions of the first kind. \footnote{The asymptotic expression of Hankel function of first kind is given by $H^{(1)}_\n(z) \sim \sq{\f{2}{\pi z}} \exp \le[i \le(z - \f{\n \pi}{2} - \f{\pi}{4}\ri)\ri]$}
Eq. (\ref{eq:h1_dS}) and (\ref{eq:h2_dS}) essentially represent the well known Bunch-Davies vacuum state \cite{Bunch-Davies-1978}.
Note that, this is a state of infinite adiabatic order however can be obtained by the zeroth order adiabatic matching as the zeroth order WKB approximation becomes exact in the limit $\eta \rightarrow -\infty$ where the matching is done. 

\subsubsection{Particle creation}

The number density of particles created during the de Sitter expansion, from Eq. \ref{eq:beta1}, is given by
\be
|\b^{dS}_k(\eta)|^2 = \f{\pi z}{4} e^{\pi \m} \le| \,f_1 H^{(1)}_{\n_-}(z) - i f_2H^{(1)}_{\n_+}(z)\ri|^2 \la{eq:beta_dS}
\ee
where $f_1$ and $f_2$ are given by Eq. (\ref{eq:f1f2e}). Note that for $m\ra 0$, the identity  $H^{(1)}_{-\n}= e^{i\n\pi}H^{(1)}_{\n}$ implies massless particles are not created which is expected and provides a nice cross-check for Eq. (\ref{eq:beta_dS}).

Then the total particle number density of specific mass $m$, summed over all modes, is given by
\be
N_\m = \f{1}{(2\pi a)^3}\int_0^\infty d^3k|\b^{dS}_k(\eta)|^2 
\ee
Changing the integrating variable from $k$ to $z=-k\eta$ leads to
\be
N_\m = \f{H^3}{2\pi^2}\int_0^\infty dz\,z^2|\b^{dS}(z)|^2. \la{eq:N_m_dS}
\ee
Eq. (\ref{eq:N_m_dS}) matches exactly with results reported in \cite{Landete:2013lpa}. 
\begin{figure}[!h]
\includegraphics[scale=.5]{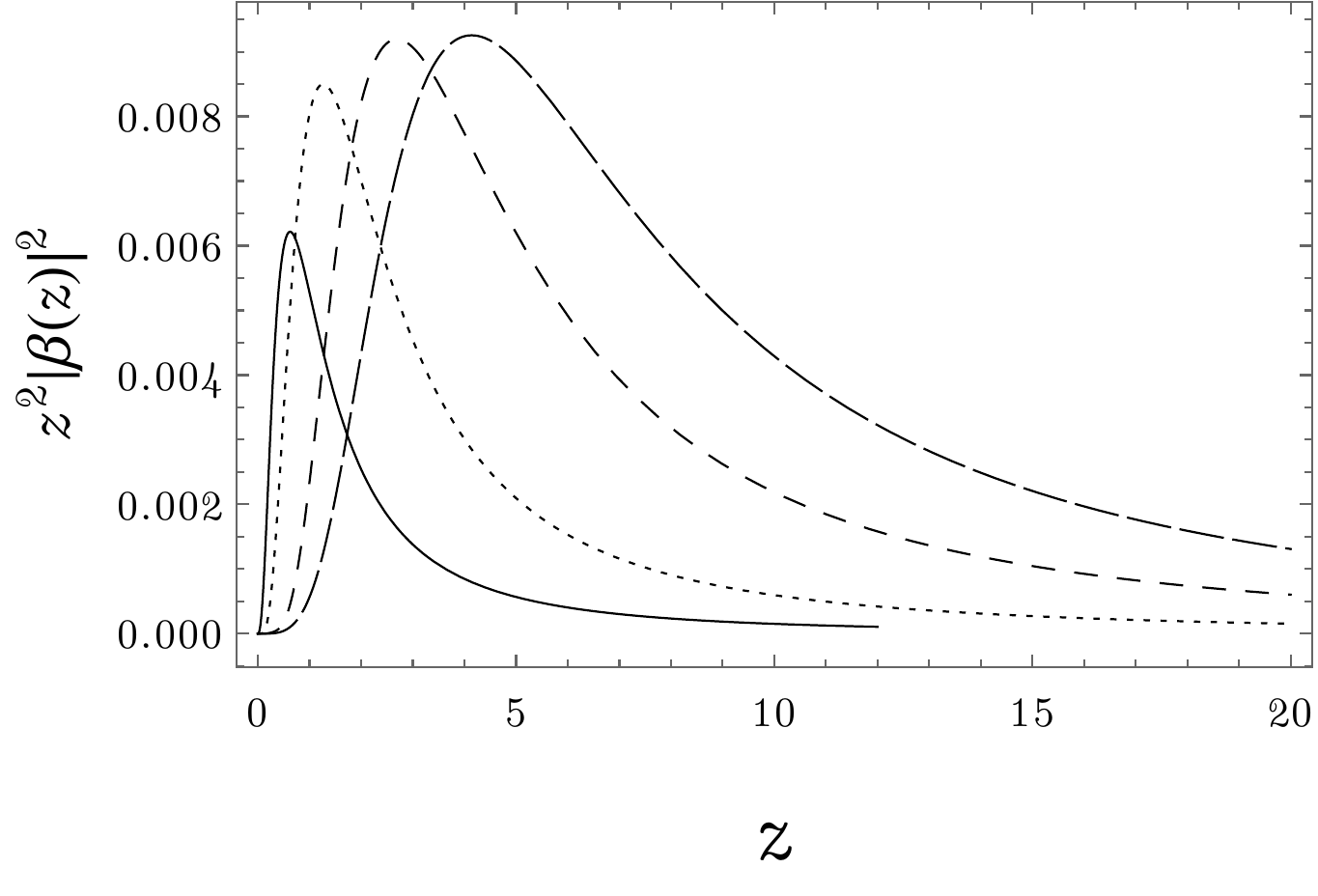} 
\caption{Variation of the integrand in Eq. (\ref{eq:N_m_dS}) (for $N_\m$) vs $z$ for masses $\m=0.5$ (continuous curve), $\m=1$ (dotted curve), $\m=2$ (small dashes) and $\m=3$ (large dashes) with $H=1$.} \la{fig:IntNvszdS}
\end{figure}
Fig. \ref{fig:IntNvszdS} shows that the integrand in Eq. (\ref{eq:N_m_dS})  peaks near $z\sim \m$ (which follows from generic properties of Hankel functions) and decays rapidly for large $z$ (which is expected as the integrand varies as $1/z^2$ with $z\ra \infty$). Further the peak value saturates to $\sim 10^{-2}$. Thus one can easily perform the integration numerically (with a cut-off at, say, $z = 10m$) and result is shown in Fig. \ref{fig:NvsmDS} that matches with \cite{delRio:2014cha}. 
Note that Fig. \ref{fig:IntNvszdS} can be looked at in two different ways. Either by fixing $\eta$ and looking at various values of $k$ or by fixing $k$ and looking at different $\eta$ where the plots need to be read from right to left.

\begin{figure}[!h]
\includegraphics[scale=.55]{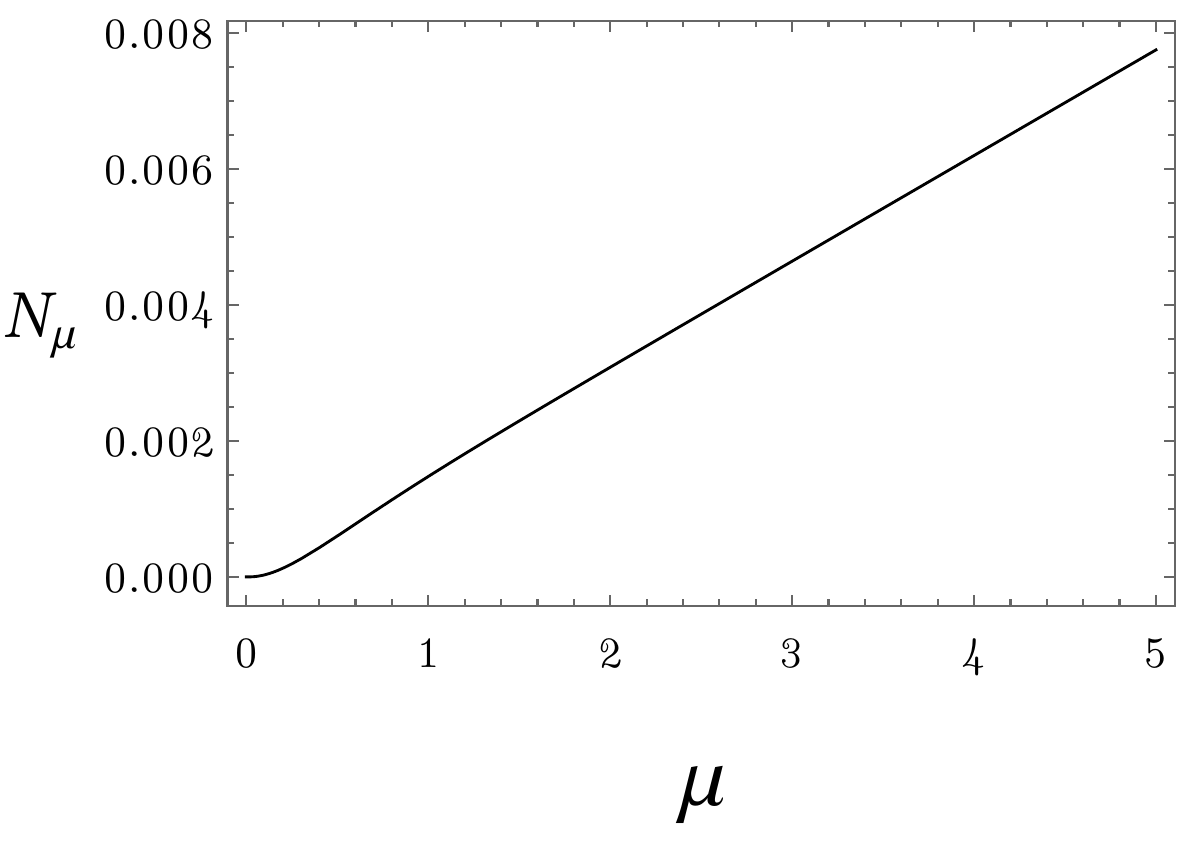} 
\caption{Total particle number density ($N_\m$) vs mass ($\m$) after summed over modes (for $0<z<10m$, $H=1$ or $\m=m$). } \la{fig:NvsmDS} 
\end{figure}


One can also derive $|\b^{dS}_k(\eta)|^2$ or $s_k(\eta)$ by solving (however numerically) the set of equations (\ref{eq:s}-\ref{eq:t}) as follows. By changing the variables from $\eta$ to $z=-k\eta$, one obtains for de Sitter expansion:
\ba
s'(z) &=& -\f{\m}{2(z^2+\m^2)} u(z), \la{eq:s_dS}\\
u'(z) &=& \f{\m}{z^2+\m^2}(1-2s(z)) - \f{2\sq{z^2+\m^2}}{z} \t(z), \la{eq:u_dS}\\
\t'(z) &=&  \f{2\sq{z^2+\m^2}}{z} u(z). \la{eq:t_dS}
\ea
with the boundary conditions $s(\infty)=u(\infty)=\t(\infty)=0$ and prime denotes derivative wrt $z$. Solving set of equations (\ref{eq:s_dS}-\ref{eq:t_dS}) leads to exactly same particle number density as given by Eq. (\ref{eq:beta_dS}) and Fig. \ref{fig:IntNvszdS}. This cross-check also verifies the applicability of equations (\ref{eq:s}-\ref{eq:t}) in cases where an analytic expression like Eq. (\ref{eq:beta_dS}) is difficult to find.
Fig. \ref{fig:NvsmDS} clearly shows that the total number density increases linearly with the mass for $m\geq H$.


\subsubsection{Renormalized energy momentum tensor}

The energy density after subtracting the vacuum contribution is given by
\be
\r_\m = \f{H^4}{2\pi^2}\int_0^\infty dz\,z^2\sq{\m^2 + z^2}|\b^{dS}(z)|^2. \la{eq:E_m_dS}
\ee
The above integral obviously has a logarithmic divergence and as Fig. \ref{fig:IntEvszdS} suggests, increases approximately as $\m^2$ with increasing mass as both the location of peaks and peak values varies almost linearly with mass.

\begin{figure}[!h]
\includegraphics[scale=.45]{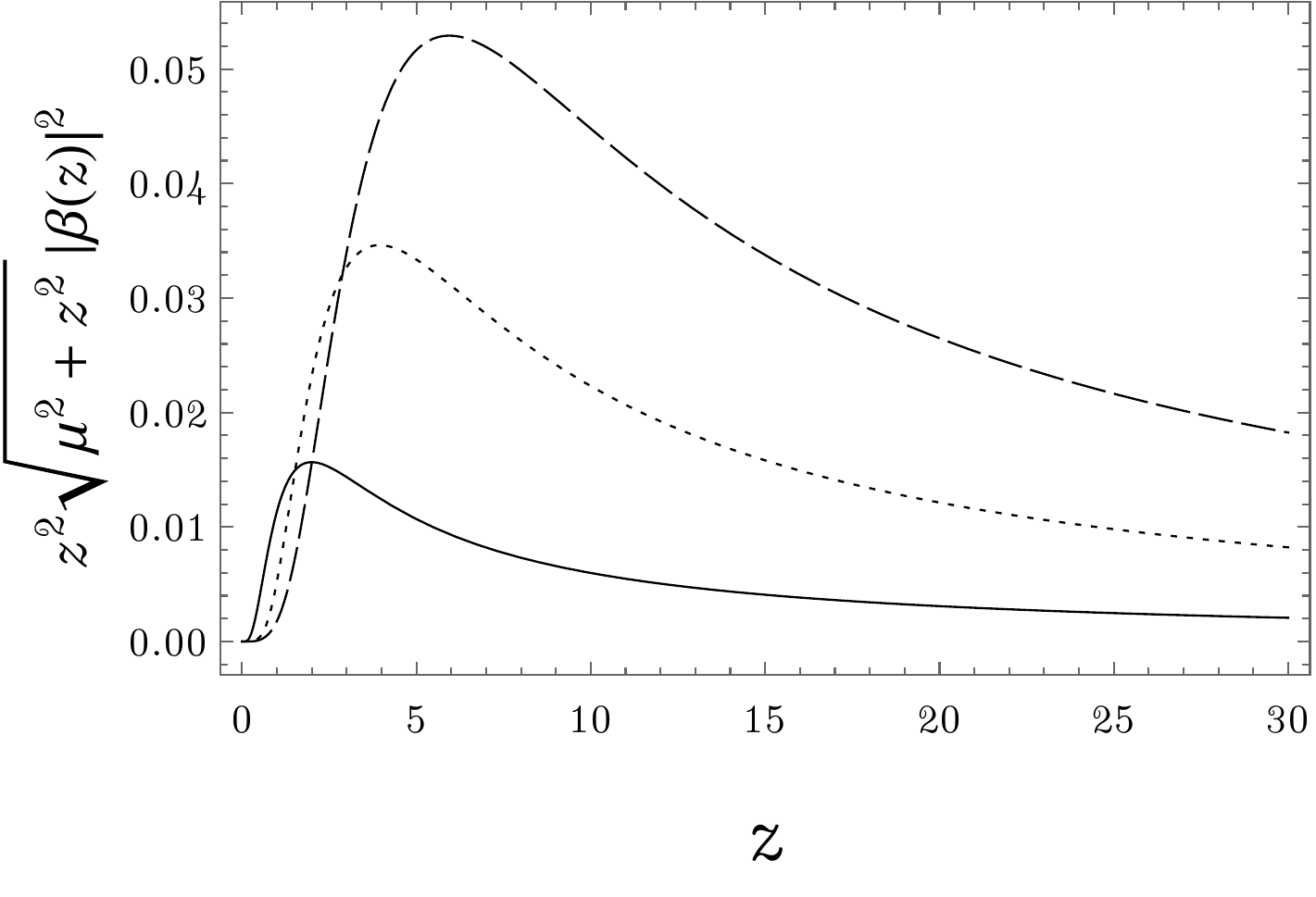} 
\caption{Variation of the integrand in Eq. (\ref{eq:E_m_dS}) (for $N_\m$) vs $z$ for masses $\m=1$ (continuous curve), $\m=2$ (dotted curve) and $\m=3$ (dashed curve)with $H=1$.} \la{fig:IntEvszdS}
\end{figure}
Let us now analyse the effect of renormalization, as formalized in the previous section, on the divergent quantities.
In de Sitter spacetime, the renormalized energy momentum tensor corresponding to de Sitter invariant states can be written in terms of its trace as
\ba
\< T_{\m\n} \> _{Ren} &=& \f{1}{4}g_{\m\n} \< T^\a_\a \> _{Ren} \\
&=& \f{1}{4}g_{\m\n} \le(\< T^\a_\a \> - \< T^\a_\a \> _{div}\ri)
\ea
where $\< T^\r_\r \> _{div}$ is comprised of the divergent quantities in the trace. Thus we need only to look into the renormalization of the trace that is given by,
\ba
\< T^\a_\a \> &=& -\f{2m}{(2\pi a)^3} \int d^3k \Big(|h^{I}_k(\eta)|^2 - |h^{II}_k(\eta)|^2\Big) \la{eq:<Tmm>dS} \\
&=& -\f{\m H^4 e^{\pi\m}}{4\pi} \int dz z^3 \Big(|H_{\n_+}^{(1)}(z)|^2 - |H_{\n_-}^{(1)}(z)|^2\Big) \la{eq:<Tmm>dSz}
\ea

As we have discussed earlier, this quantity is UV divergent and one needs to subtract upto the fourth order terms in the adiabatic expansion to regularize the energy momentum tensor. The quantity to be subtracted in case of de Sitter universe is the following,
\ba
\< T^\a_\a \> _{div} &=& -\f{2m}{(2\pi a)^3} \int d^3k \le[\f{ma}{\O_k}\le(1-2\le(s_k^{(2)} + s_k^{(4)}\ri)\ri)\ri. \nn \\ && \le. - \f{k}{\O_k}\le(u_k^{(2)} + u_k^{(4)}\ri)\ri] \la{eq:<Tmm>dSdiv} \\
&=& -\f{\m H^4}{\pi^2} \int dz \le[\f{\m z^2}{\sq{z^2 + \m^2}}\le(1-2\le(s^{(2)} + s^{(4)}\ri)\ri)\ri. \nn \\ && \le. - \f{z^3}{\sq{z^2 + \m^2}}\le(u^{(2)} + u^{(4)}\ri)\ri] \la{eq:<Tmm>dSdivz}
\ea
where the zeroth order term is essentially the vacuum contribution and $s^{(2)}(z), s^{(4)}(z), u^{(2)}(z), u^{(4)}(z)$ are given by (Appendix \ref{app-2})
\be
s^{(2)}(z) = \f{\m^2 z^2}{16(z^2 + \m^2)^3} \la{eq:s2dS}
\ee
\be
s^{(4)}(z) = -\f{\m^2 z^2(32z^4 - 69\m^2 z^2 + 4\m^2)}{256(z^2 + \m^2)^3} \la{eq:s4dS}
\ee
\be
u^{(2)}(z) = \f{mz(2z^3 - \m^2)}{4(z^2 + \m^2)^3} \la{eq:u2dS}
\ee
\be
u^{(4)}(z) = \f{-\m z(48z^6 - 186\m^2 z^4 + 79\m^2 z^2 - 2\m^6)}{32(z^2 + \m^2)^6} \la{eq:u4dS}
\ee
Using equations (\ref{eq:s2dS}-\ref{eq:u4dS}), one can easily perform the integration in Eq. (\ref{eq:<Tmm>dSdivz}) numerically and subtract the output from Eq. (\ref{eq:<Tmm>dSz}) to find $\< T^\r_\r \> _{Ren}$. Alternatively, one can use an auxiliary regulator to perform this integrations analytically as described in \cite{Landete:2013lpa} and reach the exact same result as depicted in the following.

\begin{figure}[!h]
\includegraphics[scale=.55]{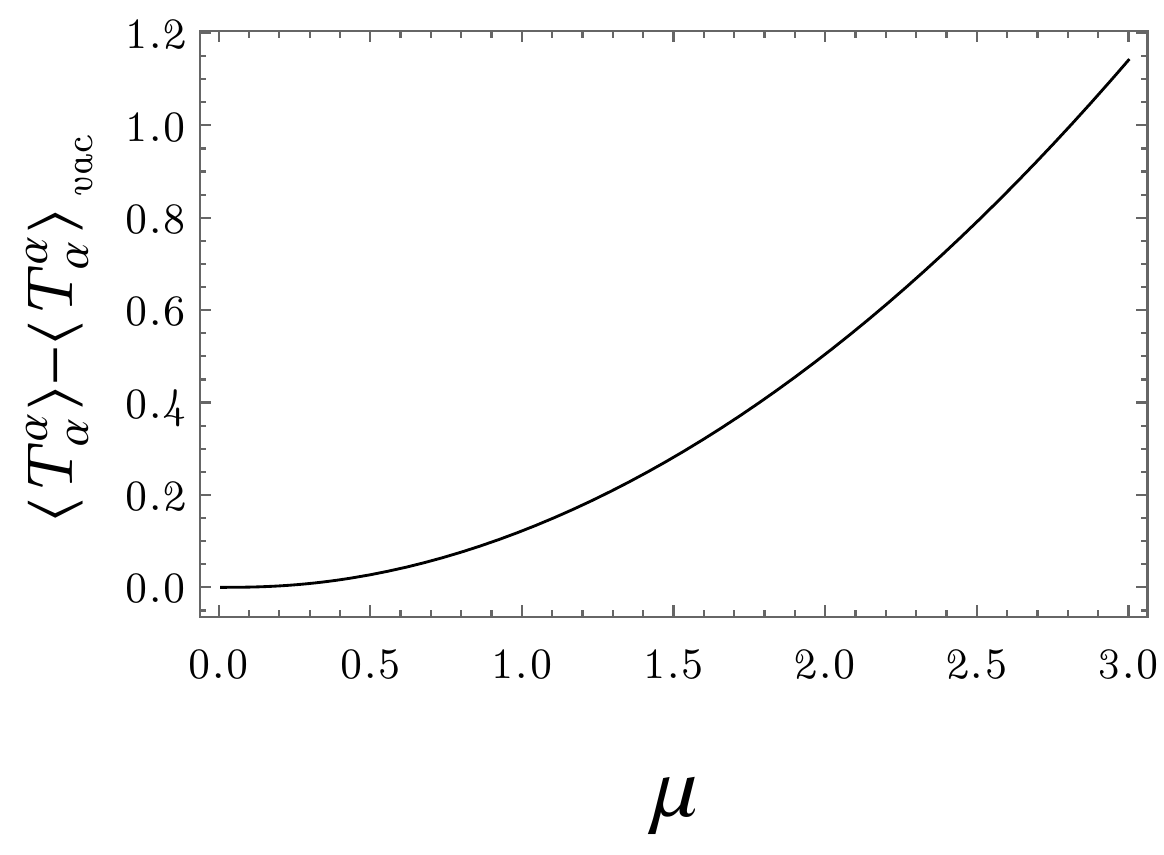} 
\caption{Trace of the stress tensor subtracted by vacuum contribution vs mass (for $0<z<30\m$ and $H=1$).} \la{fig:Trace-vac-dS}
\end{figure}

Fig. \ref{fig:Trace-vac-dS} shows that the trace after subtracting the vacuum contribution (computed using a cut-off at $z=30\m$),  increases indefinitely (almost as $\m^2$ as we have estimated) with increasing mass. Thus if massive modes get excited back reaction becomes significant and the background geometry breaks down. Remarkably, renormalization procedure solves these issue.
The fully renormalized $\< T^\r_\r \> _{Ren}$ vs. mass plot is given in Fig. \ref{fig:RenTrace-dS} and $\< T^\a_\a \> _{Ren} \ra 0$ with increasing mass. 

\begin{figure}[!h]
\includegraphics[scale=.55]{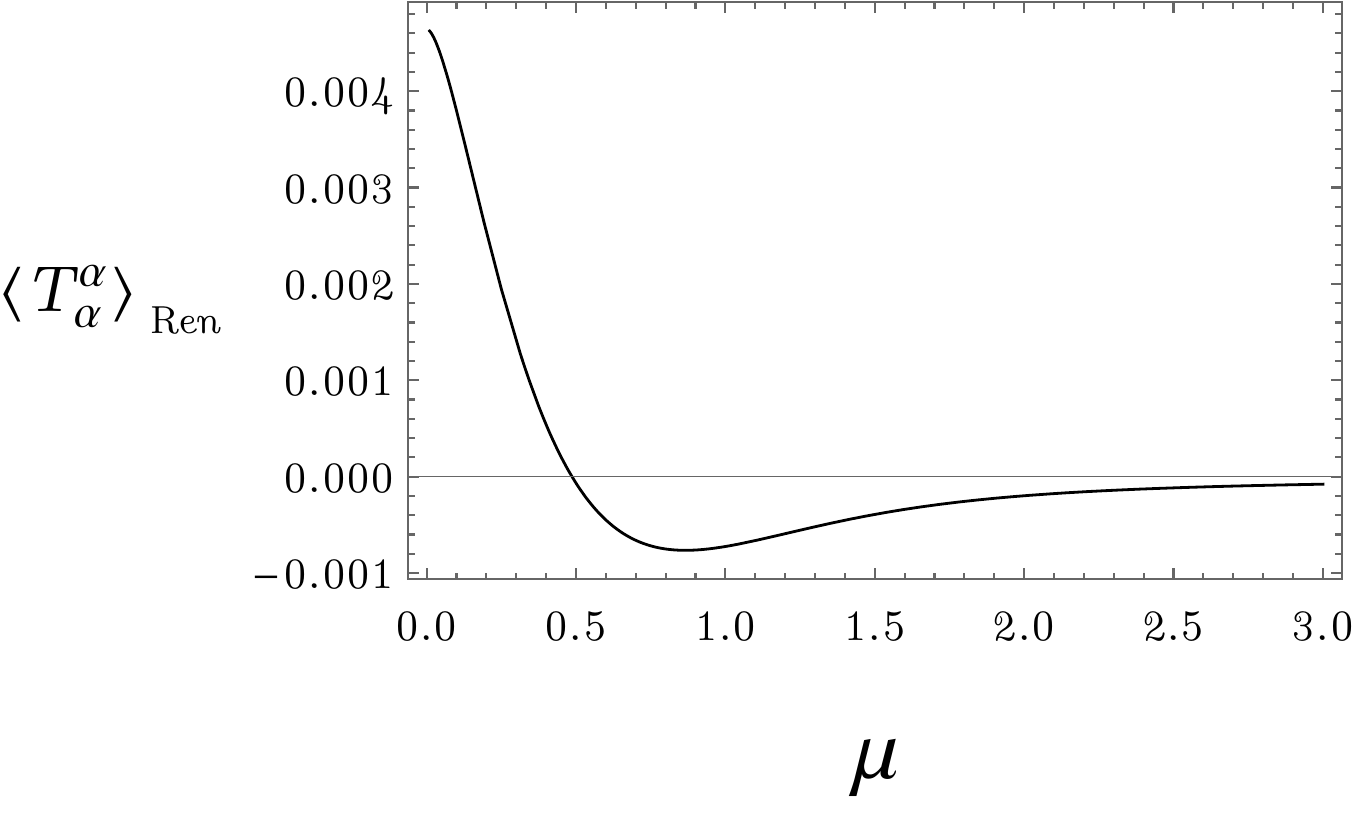} 
\caption{Renormalized trace of the stress tensor vs mass (for $0<z<10\m$ and $H=1$).} \la{fig:RenTrace-dS}
\end{figure}
Difference between Fig. \ref{fig:RenTrace-dS} and Fig. \ref{fig:Trace-vac-dS} clearly demonstrates the effect of subtracting second and fourth order adiabatic terms.  
Note that, the exact value of the fourth order subtraction term in $\< T^\r_\r \> _{div}$, which is essentially the trace anomaly with the opposite sign, is given by
\be
\< T^\a_\a \> _{div}^{(4)} = \f{\m H^4}{\pi^2} \int_0^\infty dz z^2 \le[\f{2\m s^{(4)} + z u^{(4)}}{\sq{z^2 + \m^2}} \ri] = -\f{11H^4}{240\pi^2}, \la{eq:<Tmm>dSdiv4z}
\ee
matches exactly with the value of $\< T^\a_\a \> _{Ren}$ as $m\ra 0$ (for $H=1$) in Fig. \ref{fig:RenTrace-dS}, as expected. 


\subsection{Radiation dominated universe}

The radiation dominated expansion of the universe is characterized by the cosmological scale factor $a(t) = \sq{t/t_0}$. We shall assume that the adiabatic matching is done at some $t=t_0$ when the magnitude of the scale factor is unity and $1 < t/t_0 < \infty$. Thus the conformal time is $\eta = \sq{t_0t}/2$ and the scale factor can be written as $a(\eta)=\eta/\eta_0$ where $\eta_0 =t_0/2$. Without loss of generality we will assume $t_0=2$, implying $\eta_0=1$ (i.e. $\eta_0$ to be the unit of time) and $1 < \eta < \infty$ for later numerical computations.
Solutions to Eq. (\ref{eq:h1}) and (\ref{eq:h2}), in a radiation dominated universe that has the correct asymptotic behaviour can be written in terms of parabolic cylinder functions \cite{delRio:2014bpa, Barut:1987mk}. However, as normalization of cylinder functions is difficult to achieve, the corresponding analytical investigation can only be done approximately (see \cite{delRio:2014bpa} for a detailed analysis of the asymptotic behaviour).
Note that the difficulties arise in the analysis of a radiation dominated universe essentially stems from absence of the symmetries that, for example, a de Sitter model enjoy. 
However, as we mentioned earlier, we can bypass the derivation of exact solutions and directly determine the number density, energy density and other quantities using Eqs. (\ref{eq:s}-\ref{eq:t}). Below, we carry out approximate analytic and numerical estimation to determine these quantities of physical interest.

\subsubsection{Particle creation}

The number density of particles created during the radiation dominated expansion is depicted Fig. \ref{fig:svstkRD} and \ref{fig:svstmRD} as functions of time and for varying momentum and mass. Here we derive the particle number density $s_k(\eta)$ by numerically solving  the set of equations (\ref{eq:s}-\ref{eq:t}) with the boundary conditions $s_k(\eta_0) = u_k(\eta_0) = \t_k(\eta_0) = 0$. 
In Fig. \ref{fig:svstkRD}, we have presented the evolution of $s_k(\eta)$ as function of time (with $\eta_0=1$, i.e. the horizontal axis can be regarded as $\eta/\eta_0$ for arbitrary $\eta_0$) for fixed mass and three different values of momenta. On the other hand, in Fig. \ref{fig:svstmRD}, we have presented the evolution of $s_k(\eta)$ as function of time for fixed momentum and three different values of masses.

\begin{figure}[!h]
\includegraphics[scale=.55]{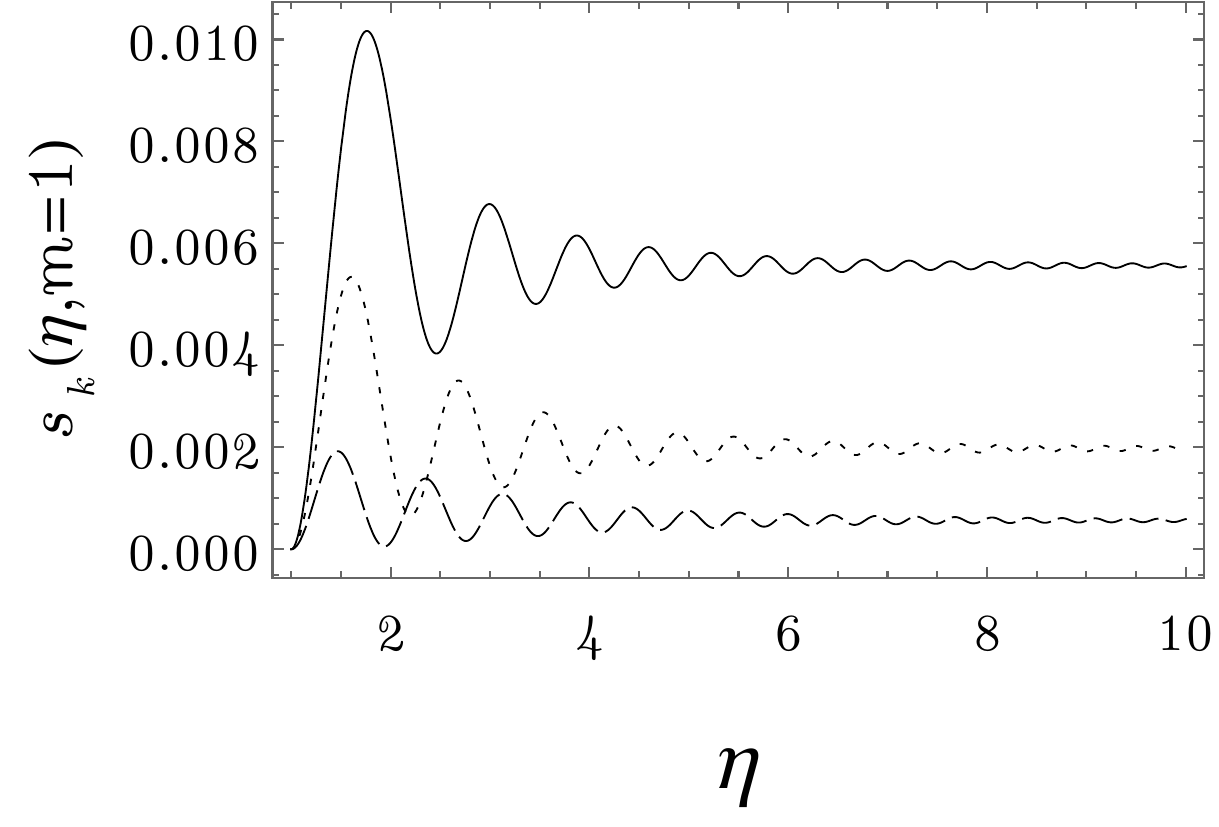} 
\caption{Particle number density vs $\eta$ for mass $m=1$ and momenta $k=1$ (continuous curve), $k=2$ (dotted curve) and $k=3$ (dashed curve) with $\eta_0=1$.} \la{fig:svstkRD}
\end{figure}
\begin{figure}[!h]

\includegraphics[scale=.58]{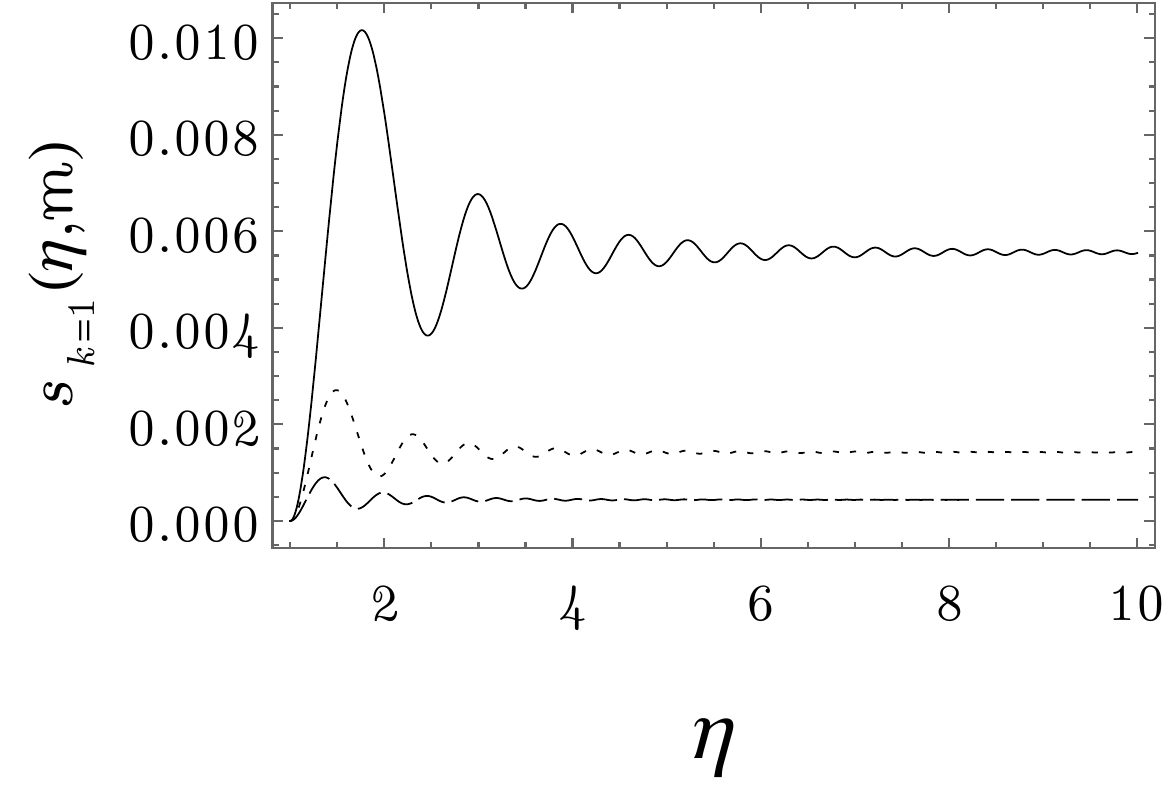} 
\caption{Particle number density vs $\eta$ for momentum $k=1$ and masses $m=1$ (continuous curve), $m=2$ (dotted curve) and $m=3$ (dashed curve) with $\eta_0=1$.} \la{fig:svstmRD}
\end{figure}

Both Fig. \ref{fig:svstkRD} and Fig. \ref{fig:svstmRD} shows that $s_k(\eta)$ grows very fast initially and after a mild oscillation quickly stabilizes to a particular value. These figures also tell us that for larger values of mass or momentum, particle creation declines sharply. Note that the oscillatory behaviour only diminishes (as $\eta \ra \eta_f$) for modes $k << m a(\eta_f)$. As the radiation dominated era ends at a  finite time, apparently, the $\eta \ra \infty$ limit is not compatible with the observed cosmological scenario. However, on physical grounds, we expect that significant contribution to the energy density of created particles should come from the modes $\O_k \sim \eta_0^{-1}$ i.e. the number of particles in modes such that $\O_k \eta_0 > 1$ is exponentially suppressed. Therefore we may assume only a few massive modes may become significantly excited.
A quantity of interest is the total particle number density in physical four volume, as given by Eq. (\ref{eq:totalPND}), of specific mass $m$, summed over all modes, leads to
\be
a^4N_m = \f{aI_m}{2\pi^2}, \la{eq:N_rad}
\ee
\be
\mbox{where}~~~ I_m = \int_0^\infty dk \,k^2 s_k(\eta), \la{eq:I_m}
\ee
which becomes constant in time as $\eta$ evolves.

\begin{figure}[!h]
\includegraphics[scale=.5]{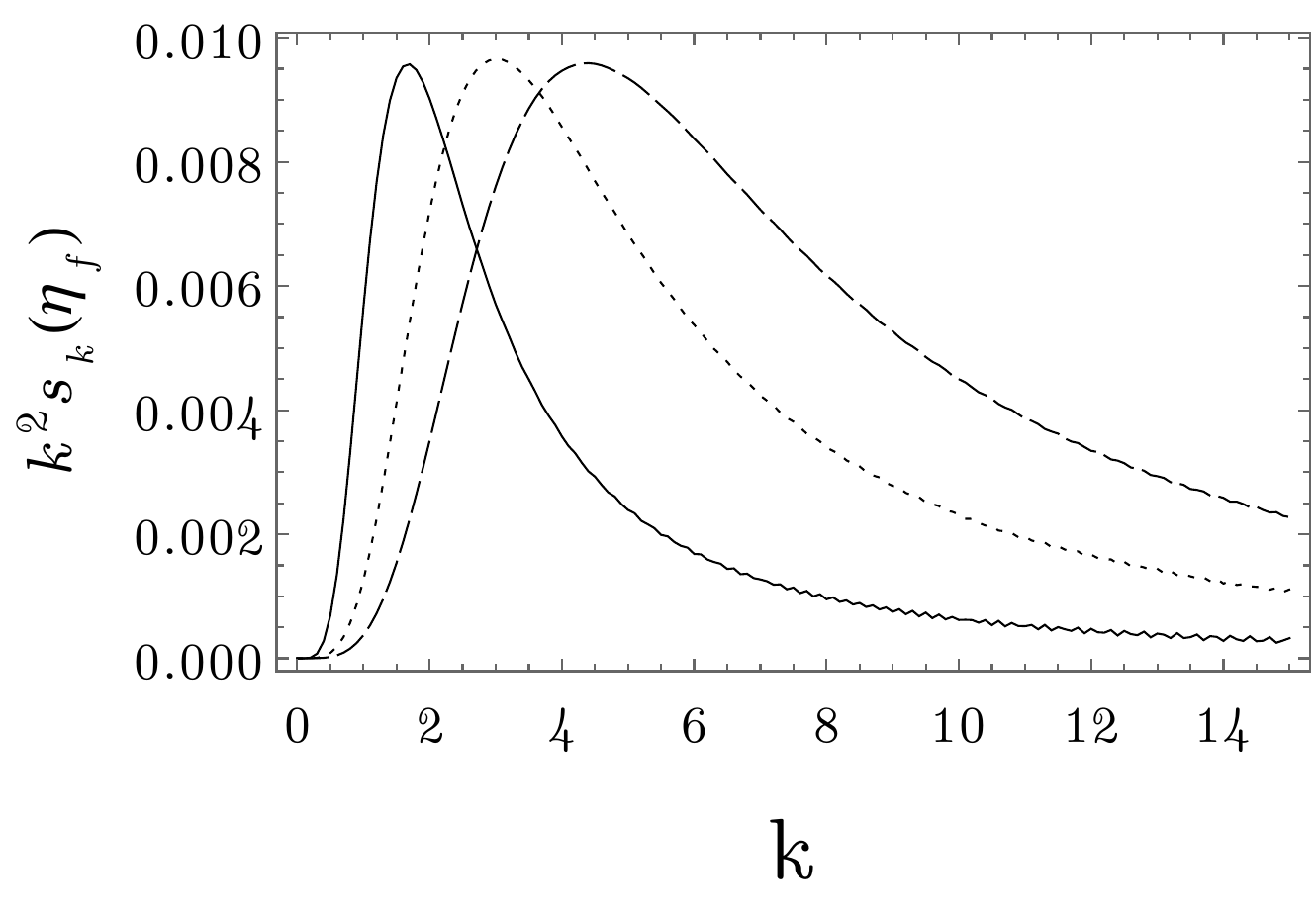} 
\caption{Variation of the integrand in Eq. (\ref{eq:I_m}) vs modes for masses $m=1$ (continuous curve), $m=2$ (dotted curve), $m=3$ (dashed curve), $\eta_f=100 \eta_0$ and $\eta_0=1$.} \la{fig:IntNvsmkRD}
\end{figure}

In Fig. \ref{fig:IntNvsmkRD}, we plot the integrand in Eq. (\ref{eq:I_m}) versus the field modes for large $\eta$ and different masses in order to estimate the integral.
The integrand peaks near $k_{max}\sim 2m$, falls off sharply (as $k^{-2}$) for approximately $k>10m$ and the peak value remains nearly same for different masses and is of the order of $10^{-2}$. As the location of the peak increases almost linearly with mass, one can approximate the area under the curves for each $k$ with a rectangle whose height $\sim 10^{-2}$ and base is proportional to mass. Thus we have, approximately, $a^4N_m \sim m$.
This relation can also be understood as follows.
Eq. (\ref{eq:N_rad}) can be written as
\ba
a^4N_m &=& \f{a}{2\pi^2} \int_0^\infty dk \,k^2 \le(s_k^{(2)} + s_k^{(4)} + s_k^{(6)} + ...\ri), \\
&=& \f{0.00186m}{\eta_0^2} - \f{0.00023}{\eta^4 m} + \f{0.00078 \eta_0^2}{\eta^8 m^3} 
- {\cal O}(m^{-5}) \la{eq:N_rad_ad}
\ea
where (see Appendix \ref{app-2}))
\be
s_k^{(2)}(\eta) = \f{m^2 k^2 \eta_0^4}{16(k^2\eta_0^2 + m^2\eta^2)^3} \la{eq:s2rad}
\ee
\be
s_k^{(4)}(\eta) = \f{21 m^4 k^2\eta_0^6(k^2\eta_0^2 - 4m^2 \eta^2)}{256(k^2\eta_0^2 + m^2\eta^2)^6} \la{eq:s4rad}
\ee
\be
s_k^{(6)}(\eta) = \f{11 m^6 k^2\eta_0^8(79k^4\eta_0^4 - 1116m^2k^2\eta_0^2 \eta^2 + 1080m^4\eta^4)}{2048(k^2\eta_0^2 + m^2\eta^2)^9} \la{eq:s4rad}
\ee
Note that, in this case, in $k \ra \infty$ limit, we have
\be
s_k^{(2)} \sim k^{-4},~~ s_k^{(4)} \sim k^{-8},~~ s_k^{(6)} \sim k^{-12}
\la{eq:asym_rad}.
\ee
Eq. (\ref{eq:N_rad_ad}), shows that the particle number density in a physical volume, for large mass, is proportional to mass and for specific mass it becomes constant in the adiabatic limit (for $\eta>>\eta_0$). 



\subsubsection{Energy density and renormalization}

The particle number depends on the method of computation because of lack of an exact notion of {\em particles} \cite{BD}. The spectrum of these particles is also found to be not thermal. Thus, one can gain a better insight by looking at the local quantities like the comoving energy density and comparing it with the critical energy density ($\r_c$) of a FLRW radiation dominated universe \cite{Weinberg} given by,
\be
a^4\r_c = \f{3\dot{a}^2}{8\pi \eta_p^2} = \f{3}{8\pi\eta_0^2\eta_p^2}
\ee
where we put $\eta_p$ as the Planck time (which is unity in natural units) for dimensional consideration. 

The energy density of created particles as given by Eq. (\ref{eq:<T00>}) (after subtracting the vacuum contribution) is
\ba
a^4\r &=& \f{2}{\pi^2}\int_0^\infty dk \,k^2 \O_k(\eta) s_k(\eta). \la{eq:rho_rad}
\ea

\begin{figure}[!h]
\includegraphics[scale=.53]{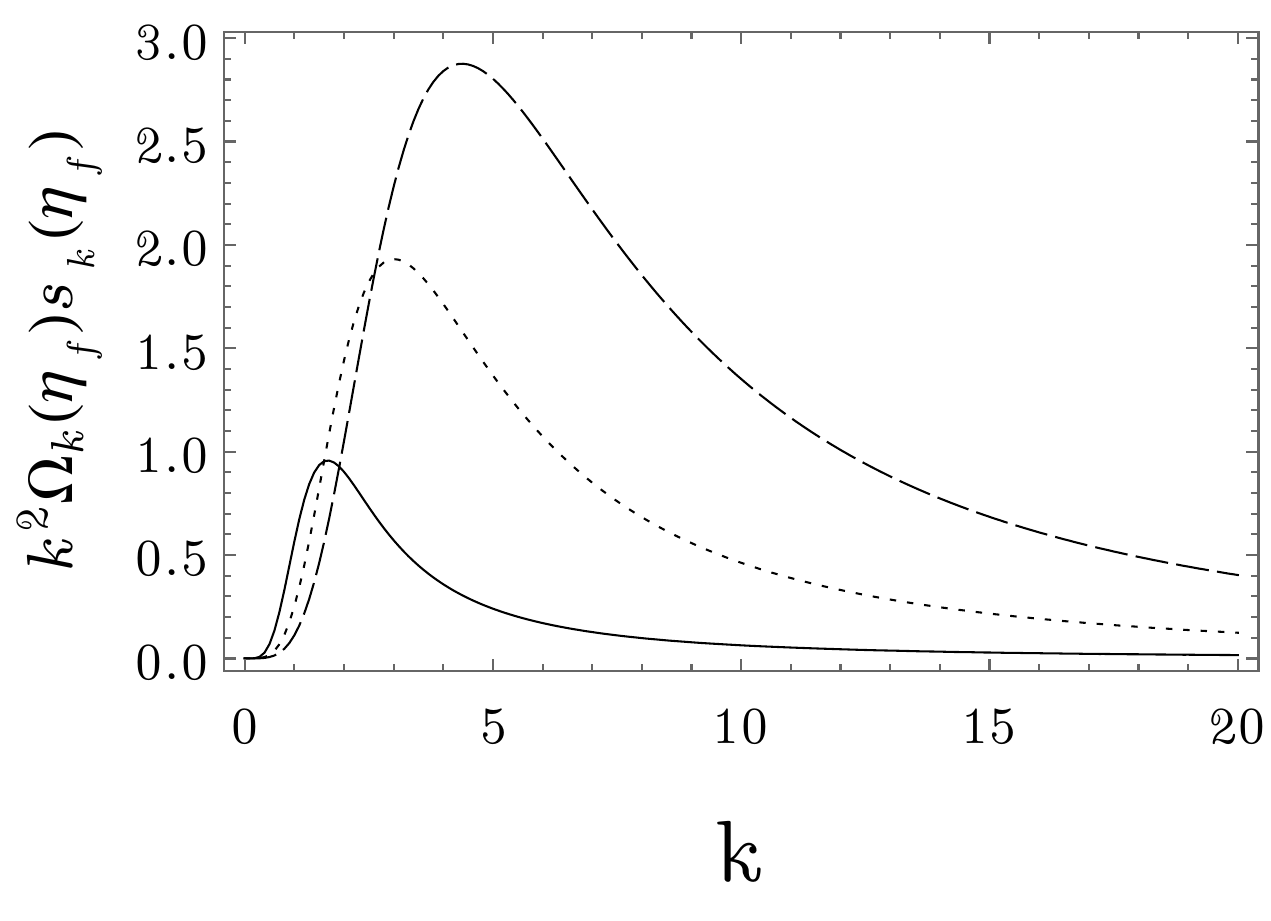} 
\caption{Variation of the integrand in Eq. (\ref{eq:rho_rad}) vs modes for masses $m=1$ (continuous curve), $m=2$ (dotted curve), $m=3$ (dashed curve), $\eta_f=100 \eta_0$ and $\eta_0=1$.} \la{fig:IntEvskmRD}
\end{figure}

Fig. \ref{fig:IntEvskmRD} shows that both the height and the location of the peak of the integrand in Eq. (\ref{eq:rho_rad}) increases almost linearly with mass and thus the integral should vary as $m^2$. However, the integral will not converge, as for large $k$, $\O_k \sim k$ and $s_k \sim k^{-4}$. To make it converge, one has to carry out the regularization procedure that should also take away the $m^2$ dependence like in the de Sitter case. 

\begin{figure}[!h]
\includegraphics[scale=.45]{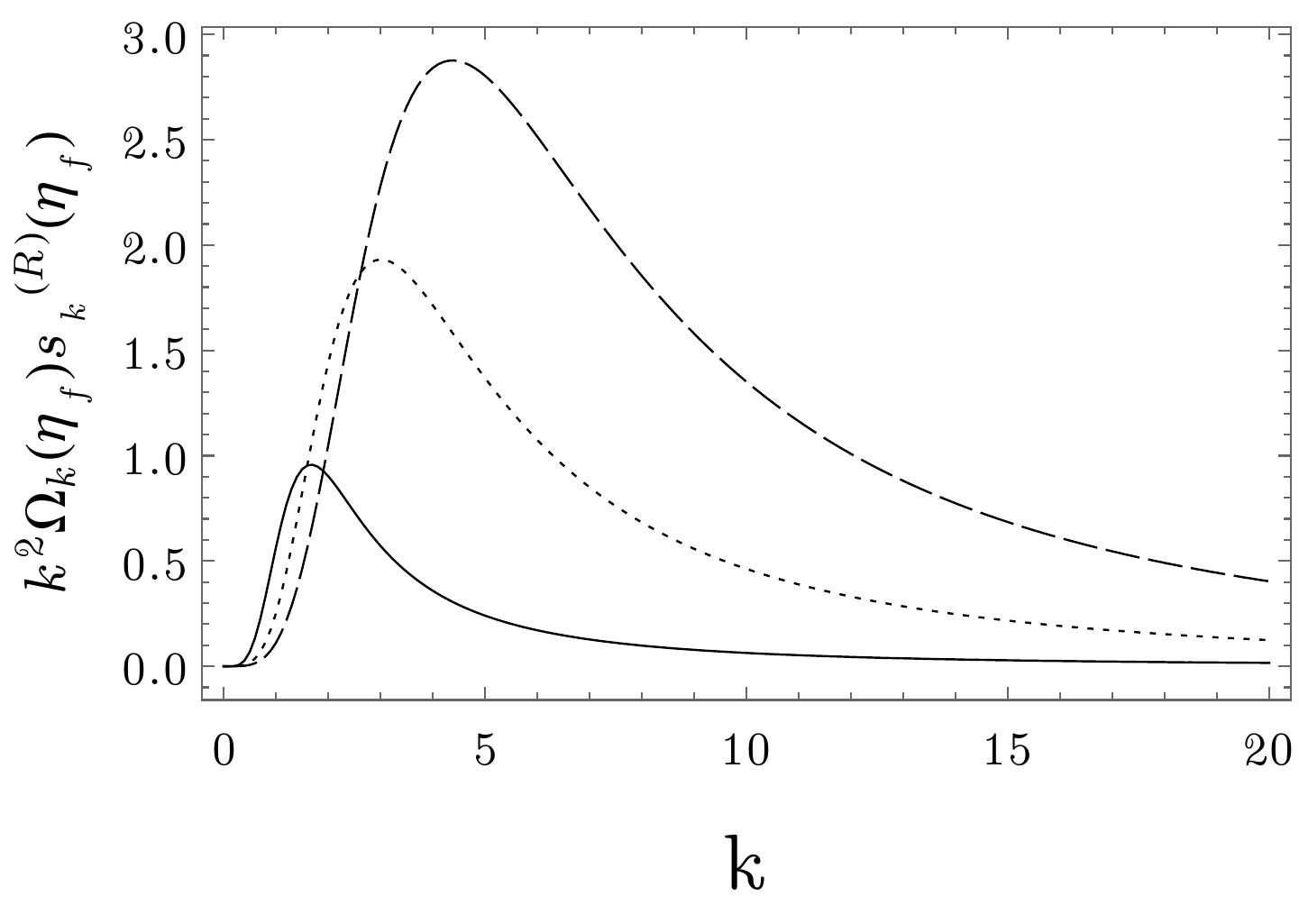} 
\caption{Variation of the renormalized version of the integrand in Eq. (\ref{eq:rho_rad}) vs modes for masses $m=1$ (continuous curve), $m=2$ (dotted curve), $m=3$ (dashed curve), $\eta_f=100 \eta_0$ and $\eta_0=1$. Here $s_k^{(R)}(\eta_f)= s_k(\eta_f) - s_k^{(2)}(\eta_f) - s_k^{(4)}(\eta_f)$.} \la{fig:IntREvskmRD}
\end{figure}
However, Fig. \ref{fig:IntREvskmRD} shows that the integrand of the renormalized energy density, as given by Eq. (\ref{eq:r_k_ren_gen}), does not change appreciably from Fig. \ref{fig:IntEvskmRD} unlike the de Sitter case. This implies that the subtraction process did not work appropriately. The reason is the following. In $s_k^{(R)} = s_k - s_k^{(2)} - s_k^{(4)}$, numerically determined $s_k(\eta)$ (presented in \ref{fig:svstkRD} and Fig. \ref{fig:svstmRD}) has become constant at very early time whereas $s_k^{(2)}(\eta)$ and $s_k^{(4)}(\eta)$ decreases with increasing $\eta$. At large $\eta_f$, the subtraction terms become negligible. We need to fix $\eta_f$ at the moment when particle creation freezes (i.e. near $\eta_0$) i.e. the regularization is needed at early times or for large modes as such. However, Fig. \ref{fig:svstkRD} and Fig. \ref{fig:svstmRD} shows that these moments are different for each mode and we could not fix $\eta_f$ consistently for numerical subtraction.  
Thus, the subtraction scheme does not work consistently unlike the de Sitter case. 
However, the subtraction method works if one can construct a fourth or higher order adiabatic state. 

Note that, we can compute analytically the renormalized energy density to the leading order as follows,
\ba
a^4\r_{_{Ren}} &=& \f{2}{\pi^2}\int_0^\infty dk \,k^2 \O_k \le(s_k - s_k^{(2)} - s_k^{(4)}\ri)\\
&=&  \f{2}{\pi^2}\int_0^\infty dk \,k^2 \O_k \le(s_k^{(6)} + s_k^{(8)} + ... \ri)\\
&=& \f{137 \eta_0^2}{4032\pi^2m^2\eta^8} + {\cal O}(m^{-4}\eta^{-12}). \la{eq:rho_rad_ren}
\ea
This gives,
\be
\f{\r_{_{Ren}}}{\r_c} = 0.0293\f{\eta_0^4\eta_p^2}{m^2\eta^8}. \la{eq:ratio_rad1}
\ee
For $\eta_p \leq \eta_0$, we have, 
\be
\f{\r_{_{Ren}}}{\r_c} \leq 0.0293\f{\eta_0^6}{m^2\eta^8}. \la{eq:ratio_rad2}
\ee
Thus renormalized energy density always remains less than the critical density and the classical background do not break down as back reaction is negligible.


\section{Summary}

We have constructed a simple formalism, within the framework of adiabatic regularization \cite{Parker-thesis, Zeldovich:1971mw, Parker:1974qw}, to derive the number density and the renormalized energy momentum tensor of spin $1/2$ particles created during the evolution of spatially flat FLRW universe. This work gives us an alternative to \cite{Landete:2013axa} and an appropriate extension of standard techniques originally introduced for scalar fields, applicable to fermions in curved space. An appropriate WKB ansatz is introduced for the `in' vacuum that satisfies the normalisation and the Wronskian condition in the adiabatic limit. The {\em non-adiabaticity} is incorporated in the time-dependent Bogolubov coefficients. 
Stress tensor components are expressed as simple linear combinations of three real and independent variables $s_k, u_k, \t_k$ which are defined using the Bogolubov coefficients. 
It is straightforward to find the adiabatic expansion of these variables in inverse powers of momenta. Subtracting terms upto fourth adiabatic order from the expansion of $\< T_{\m\n} \> $, renormalization is achieved. The renormalized quantities by construction obey the conservation law. The conformal and axial anomalies thus found are in exact agreement with those obtained from other renormalization methods. 
We have applied the formalism to two FLRW cosmological scenarios, namely de Sitter and radiation dominated universes. In the first case, we have compared the analytic and numerical results and they match. They also agree with the known results. In the second case, we have made order of magnitude estimations for the number and energy densities to the leading order. In general, renormalization takes care of the back reaction problem. In future, we plan to apply this formalism in other cosmological scenarios of interest e.g. during an anisotropic collapse \cite{Zeldovich:1971mw} or in a warped cosmological braneworld \cite{Ghosh:2008vc}.


\section*{Acknowledgments}

The author thanks A. del Rio and J. Navarro-Salas for helpful discussions and correspondence.

\appendix

\section{WKB solution}\la{app-1}
To find the WKB solution to 
\ba
{\ddot h}^I_k + \big[\O_k^2(\eta) + i Q(\eta)\big] h^{I}_{k} &=& 0, \la{eq:h1app} \\
{\ddot h}^{II}_k + \big[\O_k^2(\eta) - i Q(\eta)\big] h^{II}_{k} &=& 0, \la{eq:h2app}
\ea
let us assume
\be
h^{I}_k(\eta) \sim \exp\le[\int\Big(X(\eta) + i Y(\eta)\Big)d\eta\ri] \la{eq:wkbh1}
\ee
\be
\mbox{where}~~ X(\eta) = \f{1}{\hbar} \sum_{n=0}^{\infty}\hbar^n X_n(\eta), ~~Y(\eta) = \f{1}{\hbar} \sum_{n=0}^{\infty}\hbar^n Y_n(\eta). \la{eq:XY}
\ee
Putting Eq. (\ref{eq:wkbh1}) in Eq. (\ref{eq:h1app}) and equating the terms of zeroth order in $n$ we get
\ba
X_0^2 - Y_0^2 + \O_k^2 &=& 0, \la{eq:wkb11}\\ 
2X_0Y_0 + Q &=& 0. \la{eq:wkb12}
\ea
Similarly, solving for the first order in $n$, leads to
\ba 
\dot{X}_0 + 2X_0X_1 - 2Y_0Y_1 &=& 0, \la{eq:wkb21}\\ 
\dot{Y}_0 + 2X_0Y_1 - 2Y_0X_1 &=& 0. \la{eq:wkb22}
\ea
Higher order terms can be neglected in the adiabatic approximation. Solving Eqs (\ref{eq:wkb11}) and (\ref{eq:wkb12}) we get
\be
X_0 \approx \f{Q}{2\O_k}, ~~ Y_0 \approx \O_k.
\ee
Similarly, Eqs (\ref{eq:wkb21}) and (\ref{eq:wkb22}) gives
\be
X_1 \approx -\f{\dot{\O}_k}{2\O_k}, ~~ Y_1 \approx 0.
\ee
This leads to
\be
h^{I}_k(\eta) \sim \sq{\f{\O_k + ma}{2\O_k}} \exp\le[ i\int\O_k d\eta\ri]
\ee
Similarly one can solve Eq. (\ref{eq:h2app}). Note that the approximations made above are valid in the adiabatic limit. Therefore Eq. (\ref{eq:g1g2}) represents the adiabatic vacuum.


\section{$s_k$, $u_k$ and $\t_k$ of different adiabatic orders} \la{app-2}

Terms in adiabatic expansion of $s_k$, $u_k$ and $\t_k$ can be derived solving Eqs (\ref{eq:s}-\ref{eq:t}) in the following way. For $k\ra \infty$, $s_k$, $u_k$, $\t_k$ and there temporal variations must tend to zero. Therefore, Eq. (\ref{eq:u}) for large $k$ implies,
\be
0 \sim 2F - 2\O_k \t_k, 
\ee
which further implies that the leading term in the adiabatic expansion of $\t_k$ is of adiabatic order one, i.e.
\be
\t_k^{(1)} \sim \f{F}{\O_k} = \f{mk\dot{a}}{2\O_k^3}. \la{eq:t1} 
\ee
Putting Eq. (\ref{eq:t1}) in Eq. (\ref{eq:t}) we get the leading term in the adiabatic expansion of $u_k$ which is of order two, 
\be
u_k^{(2)} \sim \f{\dot{\t}_k^{(1)}}{2\O_k} = - \f{3m^3ka\dot{a}^2}{4\O_k^6} + \f{m k \ddot{a}}{4\O_k^4}.	 \la{eq:u2} 
\ee
Similarly, by putting Eq. (\ref{eq:u2}) in Eq. (\ref{eq:s}) we get the leading term in the adiabatic expansion of $s_k$ which is again of order two, 
\be
s_k^{(2)} \sim \int F u_k^{(2)} d\eta = \f{m^2k^2\dot{a}^2}{16\O_k^6}. \la{eq:s2} 
\ee
Now putting Eq. (\ref{eq:s2}) again back in Eq. (\ref{eq:u}) we get the next to leading term in the adiabatic expansion of $\t_k$ which is of adiabatic order three. This iteration leads to the equations (\ref{eq:u^r}-\ref{eq:t^r}) that give for higher orders:
\be
\t_k^{(3)} = \f{5m^3k^3\dot{a}^3}{16\O_k^9} - \f{15m^5ka^2\dot{a}^3}{8\O_k^9} + \f{5m^3k\dot{a}\ddot{a}}{4\O_k^7} - \f{mk\dddot{a}}{8\O_k^5}, \la{eq:t3} 
\ee

\ba
u_k^{(4)} &=& \f{35m^3k^5\dot{a}^2\ddot{a}}{32\O_k^{12}} + \f{5m^7ka^5\ddot{a}^2}{8\O_k^{12}} - \f{15m^7ka^5\dot{a}\dddot{a}}{16\O_k^{12}} + \f{5m^3k^5a\ddot{a}^2}{8\O_k^{12}}  \nn \\ 
&& - \f{105m^5k^3a\dot{a}^4}{32\O_k^{12}}  + \f{15m^3k^5a\dot{a}\dddot{a}}{16\O_k^{12}} + \f{105m^7ka^3\dot{a}^4}{16\O_k^{12}} \nn \\
&&  + \f{5m^5k^3a^3\ddot{a}^2}{4\O_k^{12}} + \f{15m^5k^3a^3\dot{a}\dddot{a}}{8\O_k^{12}} - \f{mk^7\ddddot{a}}{16\O_k^{12}} \nn \\ 
&& - \f{175m^5k^3a^2\dot{a}^2\ddot{a}}{32\O_k^{12}} - \f{3m^3k^3a^2\ddddot{a}}{16\O_k^{10}} - \f{105m^7k\dot{a}^2\ddot{a}}{16\O_k^{12}}, \la{eq:u4} 
\ea

\ba
s_k^{(4)} &=& \f{m^4k^4\dot{a}^4}{16\O_k^{12}} - \f{m^6k^2a^2\dot{a}^4}{4\O_k^{12}} + \f{7m^4k^2a\dot{a}^2\ddot{a}}{32\O_k^{10}} \nn \\
&& + \f{m^2k^2\ddot{a}^2}{64\O_k^8} - \f{m^2k^2\dot{a}\dddot{a}}{32\O_k^8}. \la{eq:s4}
\ea
Higher order terms can be derived in similar way. 

%
%

\section{Useful geometric quantities}

Following are few useful formulas for FLRW geometry.
Affine connections:
\be
\Gamma^0_{00} = \Gamma^0_{ii} = \Gamma^i_{i0} = \f{\dot{a}}{a}, ~~~~ i=1,2,3.
\ee

\be
\mbox{Ricci scalar:}~~ R = 6\f{\ddot{a}}{a}, \la{eq:R}
\ee

\be
R_{\m\n}R^{\m\n} = 12\le(\f{\dot{a}^4}{a^8} - \f{\dot{a}^2\ddot{a}}{a^7}	+ \f{\ddot{a}^2}{a^6}\ri). \la{eq:Rmnsq	}
\ee

\be
\mbox{Gauss-Bonnet scalar:}~~G = 24\, \f{a\dot{a}^2\ddot{a} - \dot{a}^4}{a^8}.
\ee

\be
\square R = 6 \le(\f{3\ddot{a}^2}{a^6} - \f{6\dot{a}^2\ddot{a}}{a^7} + \f{4\dot{a}\dddot{a}}{a^6} - \f{\ddddot{a}}{a^5}\ri), \la{eq:boxR}
\ee

\end{document}